\documentclass[11pt]{article}
\usepackage{graphicx}
\usepackage{amssymb}
\usepackage{fancyhdr}
\font\bigbf=cmssbx10 scaled\magstep2

\def\AEF{A.E. Faraggi}
\def\JHEP#1#2#3{{\it JHEP}\/ {\bf #1} (#2) #3}
\def\NPB#1#2#3{{\it Nucl.\ Phys.}\/ {\bf B#1} (#2) #3}
\def\PLB#1#2#3{{\it Phys.\ Lett.}\/ {\bf B#1} (#2) #3}
\def\PRD#1#2#3{{\it Phys.\ Rev.}\/ {\bf D#1} (#2) #3}
\def\PRL#1#2#3{{\it Phys.\ Rev.\ Lett.}\/ {\bf #1} (#2) #3}
\def\PRT#1#2#3{{\it Phys.\ Rep.}\/ {\bf#1} (#2) #3}
\def\MODA#1#2#3{{\it Mod.\ Phys.\ Lett.}\/ {\bf A#1} (#2) #3}
\def\IJMP#1#2#3{{\it Int.\ J.\ Mod.\ Phys.}\/ {\bf A#1} (#2) #3}

\def\MPLA#1#2#3{{\it Mod.\ Phys.\ Lett.}\/ {\bf A#1} (#2) #3}
\def\IJMPA#1#2#3{{\it Int.\ J.\ Mod.\ Phys.}\/ {\bf A#1} (#2) #3}
\def\EPJC#1#2#3{{Eur.\ Phys.\ Jour.}\/ {C \bf #1} (#2) #3}

\def\etal{{\it et al\/}}

\fancypagestyle{plain}{ \fancyhf{} \cfoot{-~\thepage~-} }

\def\ni{\noindent}

\def\beq{\begin{equation}}
\def\eeq{\end{equation}}
\def\beqn{\begin{eqnarray}}
\def\eeqn{\end{eqnarray}}

\title{Phenomenological survey of free fermionic heterotic--string models}
\author{Alon E. Faraggi\\ \it \small  
Theoretical Physics Division, University of Liverpool, Liverpool L69 7ZL, UK
\\ \\
June  30th, 2008\\  \\
\it \small \sl Talk at the
Sixth Simons Workshop in Mathematics and Physics \\
\it \small \sl  Stony Brook University, June 16 - July 12, 2008 } 
\date{}

\begin{document}

\maketitle

{\bf~~~~~~~~~~~~~~~~~~~~~~~~~~~~~~~~~~~~~~~~~~~~~~~~ Abstract}

{\it \small  A glossary overview of the phenomenological studies of
quasi--realistic free fermionic heterotic string models is presented.
I elaborate on
the correspondence of these models with $Z_2\times Z_2$ orbifolds.}
\\

\ni {\bf Introduction}

The standard model of particle physics passes all experimental observations 
with flying colors. The gauge charges of the standard model matter states
are strongly suggestive of the embedding of the standard model in larger grand 
unified groups. This is particularly striking in the context of $SO(10)$ 
grand unification in which each of the standard model matter generation is 
embedded in a single 16 spinorial representation of $SO(10)$. We recall 
that the standard model gauge charges were experimentally discovered and 
therefore are experimental observables. To account for these charges in 
the framework of the standard model requires $3\times 3\times 6=54$
distinct parameters, taking into account the three group factors, the 
three generations and the six multiplets (including the right--handed
neutrino) within each generation. The embedding of the standard model
reduces this number to one, being the number of spinorial representations
needed to accommodate the three generations of the standard model, namely
three. The evidence for the realisation of grand unification structures in 
nature is therefore striking indeed. 

The standard model and grand unification in themselves cannot, however,
be the end of the story. The first apparent question that pops to
mind is how three generations came to be and not two, four or five. 
Next, there are the various mass and flavor mixing parameters of the
standard model. The origin of these parameters is not explained in the 
context of the standard model, nor in grand unified theories.

It is plausible therefore that to seek answers to these questions
one must explore the origins of the Standard Model at a more basic 
level. In modern parlance this means at an energy scale, which is 
above the the GUT energy scale, {\it i.e.} the Planck scale, where
the strength of the gravitational interaction is comparable to that 
of the gauge interactions. We are lead to this conclusion by the 
structure of the standard model itself. 

String theory provides a self consistent framework for the synthesis of 
quantum mechanics and gravity. It is a natural extension of point quantum 
field theories. It admits a quantised particle interpretation, which in
is a highly non--trivial result. Furthermore, the internal particle 
attributes, which in point particle gauge theories are ad hoc, arise 
in string theory from the internal consistency conditions. We can interpret
these internal degrees of freedom as extra space time dimensions.
The important feature of string theory is that while
providing a consistent approach to quantum gravity it gives
rise to the gauge and matter structures that are used in contemporary
quantum field theories and the standard model. This
enables the development of a phenomenological approach to quantum
gravity by constructing string models that aim to reproduce the 
standard model and in turn can be used to explore the dynamics
of string theory and its fundamental properties from a phenomenological 
point of view. 

The five ten dimensional string theories, as well as eleven supergravity are
believed to be limits of a more fundamental theory. Any one of this limits
can be used to construct phenomenological string models. As limits of 
a more fundamental theory we should not expect any of the limits to provide
a complete description of the true vacuum but merely to probe some
of its properties. As the standard model data
favor its embedding in $SO(10)$, the two pivotal requirements from 
a phenomenological string vacuum is the existence of three generations
and their embedding into $SO(10)$ multiplets. The perturbative string limit
that facilitates the embedding in $SO(10)$ is the heterotic--string
as it is the limit that produces spinorial representations in the 
perturbative spectrum. Thus, to preserve these two key properties
of the standard model spectrum the perturbative string
limit that should be used is the heterotic string. It is likely that
to obtain insight into other properties of the true vacuum other
limits other perturbative string limits should be used. For example,
the dilaton exhibits a run away behaviour in the perturbative heterotic
limit and its stabilisation requires moving away from that limit. 

The study of phenomenological string vacua proceeds with the
compactification of the heterotic--string from ten to four dimensions. 
A class string compactifications that preserve the $SO(10)$ embedding of
the Standard Model spectrum are those that are based on the $Z_2\times
Z_2$ orbifold and have been extensively studied by utilizing the
so--called free fermionic formulation
\cite{fsu5,fny,alr,nahe,eu,top,cslm,otherrsm,cfn,cfs,su421,fmt}.

There are of course a large number of requirements that a realistic string 
vacuum should satisfy. Here I list a few of these requirements:

\begin{itemize}
\item{ $\longrightarrow~~~SU(3)\times SU(2)\times U(1)^n\times hidden$}
\item{ Three generations}
\item{ Proton stable ~~~~~~~~($\tau_{\hbox{P}}>10^{30}$ years)}
\item{ Higgs doublets $\oplus$ potentially realistic Yukawa couplings}
\item{ N=1 SUSY~~~~~~~~~~~~(N=0)}
\item{ Agreement with $\underline{\sin^2\theta_W}$ and %
$\underline{\alpha_s}$ at $M_Z$ (+ other observables).} 
\item{ Light left--handed neutrinos}
\item{ $SU(2)\times U(1)$ breaking \\
~~~~~~~SUSY breaking\\
~~~~~~~No flavor changing neutral currents\\
~~~~~~~No strong CP violation\\
~~~~~~~Exist family mixing and weak CP violation}
\item{ +~~ \bf{...}}
\item{ +~~~~~~~~~~~~~~~~{\bigbf{GRAVITY}}}
\end{itemize}

The fermionic formulation was developed in the mid--eighties \cite{fff}.
Just as the point particle time parameter spans a world--line,
the string time and internal parameters span the two dimensional
string world--sheet.
The equivalence of bosons and fermions of a two dimensional 
conformal field theory entails that a model constructed
using the fermionic approach correspond to a model
constructed using the bosonic approach in which the
target--space is compactified on a six dimensional internal manifold.
In this vein the free fermionic formalism correspond to using
a free bosonic formalism in which the radii of the internal dimensions 
are fixed at a special point in the compact space.
Deformation from the special point in the moduli space are parametrized 
in terms of world--sheet Thirring interactions among the world--sheet 
fermions \cite{egrs}.
This equivalence is merely the simplest illustration of the
relation between world--sheet rational conformal field theories
and manifolds with $SU(n)$ holonomy \cite{gepner}.
The simplicity of the free fermionic formalism entails that
the string consistency constraints are solved in terms of
the world--sheet free fermion transformation properties on the string
world--sheet, which are encoded in sets of basis vectors and
one--loop GSO projection coefficients among the basis vectors.
The formalism to extract the physical spectrum and superpotential
interaction terms are also straightforward. The simplest free fermionic
constructions correspond to a $Z_2\times Z_2$ orbifold of a six dimensional
toroidal manifold, augmented with discrete Wilson lines that are needed
to break the $SO(10)$ GUT symmetry.
The quasi--realistic free fermionic heterotic--string standard--like models
were constructed in the late eighties and early nineties. They provide a
concrete framework to study many of the issue that pertain to the
phenomenology of the Standard Model
and string unification. A few highlights of these studies are listed below: 

\begin{itemize}
\item[{$\bullet$}] { Top quark mass $\sim$ 175--180GeV  \cite{top,top95}}
\item[{$\bullet$}] { Generation mass hierarchy \cite{fmm}}
\item[{$\bullet$}] { CKM mixing  \cite{ckm}}
\item[{$\bullet$}] { Stringy seesaw mechanism \cite{seesaw,seesawII}}
\item[{$\bullet$}] { Gauge coupling unification \cite{gcu,df}}
\item[{$\bullet$}] { Proton stability \cite{ps}}
\item[{$\bullet$}] { Squark degeneracy \cite{fp2}}
\item[{$\bullet$}] { Minimal Standard Heterotic String Model (MSHSM) 
                                                                \cite{cfn}}
\item[{$\bullet$}] { Moduli fixing \cite{modulifixing}}
\item[{$\bullet$}] { 
Classification \& spinor--vector duality \cite{classification}}
\end{itemize}

Perhaps, the most tantalising achievement is the successful calculation of the
top quark mass, which was obtained several years prior to the experimental
discovery, and in the correct mass range. This 
calculation demonstrated how string theory enables the
calculation of the fermion--scalar
Yukawa couplings in terms of the unified gauge coupling.
Furthermore, the string  models offered an explanation for the 
hierarchical mass splitting between the top and bottom quarks.
The top quark Yukawa coupling is obtained at the cubic level
of the superpotential and is of order one, whereas the Yukawa 
couplings of the lighter quarks and leptons are obtained from
nonrenormalizable operators that are suppressed relative to the 
leading cubic level term. Thus, only the top quark mass is
characterised by the electroweak scale and the masses of the lighter
quarks and leptons are naturally suppressed compared to it. 
As the heavy generation Yukawa couplings
are obtained at low orders in the superpotential, the calculation of these
Yukawa couplings is robust and is common to a large class of models. 
The analysis of fermion masses was then further pursued, and
quasi--realistic fermion mass textures were shown to arise for reasonable
choices of supersymmetric flat directions. Issues like left--handed
neutrino masses, gauge coupling unification, proton stability and squark 
degeneracy were studied in concrete quasi--realistic free fermionic string
models and for detailed solutions of the supersymmetric flat direction 
constraints. While an attempt to find a single solution that satisfies all
the variety of phenomenological requirements listed above was not pursued,
it was demonstrated that all of the above requirements can find
satisfactory solutions in the context of the free fermionic string models.
It was also demonstrated in ref. \cite{cfn} that the free fermionic
heterotic string vacua give rise to models that produce in the 
observable charged sector below the string 
unification scale solely the matter
spectrum of the minimal supersymmetric standard model.
Such models are dubbed Minimal Standard Heterotic 
String Models (MSHSM). The free fermionic models also provide important
clues to the problem of moduli fixing in string theory. They 
highlight the fact that string theory may utilize geometrical structures
that do not have a classical correspondence. Primarily, they allow boundary
conditions that distinguish between the left-- and right--moving
coordinates of the six dimensional compactified space. 
Such boundary conditions necessarily lead to the projection
of the moduli fields associated with the extra internal coordinates.
The free fermionic models have also been instrumental in recent years to
unravel a new duality symmetry under the exchange of spinor and vector 
representations of the GUT group. 

String theory predicts that the number of degrees of freedom giving 
rise to the gauge symmetries of the standard model should be augmented
by a specific number of additional degrees of freedom. An naive
interpretation of some of those is as extra space--time dimensions.
These additional degrees of freedom may be out of reach of contemporary
experiments, and the development of phenomenological string models
aims at bridging the gap. String models give rise to additional
symmetries and matter sectors that do not arise in grand unified theories. 
These include: gauge symmetries that are external to the GUT symmetries and
may play a pivotal role in explaining proton stability \cite{zp};
matter states
that arise due to the breaking of non--Abelian gauge symmetries
by Wilson lines, which gives rise to matter states that do not obey the
GUT charge quantisation, and may lead to stable string relics
\cite{ccf}; 
specific soft SUSY breaking patterns and consequently specific predictions
for the superpartners mass spectrum \cite{fp2,dedes}. 
While all of these will be 
parametrised in terms of point quantum field theory parameters, their
experimental observations will provide further evidence for the 
validity of string theory and specific string compactifications
with which they are compatible. The final step in this program 
is to seek the all elusive dynamical mechanism, based on first
principles, that singles out the string vacuum. The free fermionic
models, and the association of the free fermionic point in
the moduli space with the self--dual point under $T$--duality,
suggests that self--duality play a vital role in this 
selection principle \cite{sdvs}. 

\ni {\bf Correspondence with $Z_2\times Z_2$ orbifold}

I elaborate here on the relation to $Z_2\times Z_2$ orbifold in which there has
been some recent interest \cite{dw}. In general, due to the equivalence of two 
dimensional fermions and bosons we can anticipate that any model constructed by
using world--sheet fermions can also be constructed by using world--sheet
bosons. The constructions using world--sheet bosons are the toroidal orbifolds,
whereas the fermionic construction is formulated at the point in the moduli
space at which the fermions are free. Models in this formalism are
defined in terms of boundary condition basis vectors and one loop generalized 
GSO projection coefficients. The correspondence between free fermion
models and bosonic constructions can be illustrated by starting with the 
set of basis vectors
\beq
\{ 1,S,\xi_1,\xi_2\},
\label{neq4set}
\eeq
generates a model with $N=4$ space-time
supersymmetry and ${\rm SO}(12)\times {\rm E}_8\times {\rm E}_8$ gauge
group. The basis vector $S$ is the space--time supersymmetry generator and
the two basis vectors $\xi_1$ and $\xi_2$ produce the two spinorial 128 representation
of $SO(16)$ and enhance $SO(16)\times SO(16)$ to ${\rm E}_8\times {\rm E}_8$. 
The free fermionic realization of the six compactified dimensions 
gives rise to the maximal $SO(12)$ enhanced symmetry. The same model
can be constructed by using the bosonic construction.
The action for the D--dimensional compactified string is given by,
$$
S={1\over{8\pi}}
\int{d^2\sigma({G_{ij}\partial^\alpha{X^i}\partial_\alpha{X^j}+
\epsilon^{\alpha\beta}B_{ij}\partial_\alpha{X^i}\partial_\beta{X^j}})}~,$$
where,
$$G_{ij}={1\over2}{\sum_{I=1}^D}R_ie_i^IR_je_j^I~,$$
is the metric of the six dimensional compactified space 
and $B_{ij}=-B_{ji}$ is the antisymmetric tensor field. 
The $e^i=\{e_i^I\}$ are six linear independent vectors normalized
to $(e_i)^2=2$.
The left-- and right--moving momenta are given by,
\beq
P^I_{R,L}=[m_i-{1\over2}(B_{ij}{\pm}G_{ij})n_j]{e_i^I}^*~,
\label{lrmomenta}
\eeq
where the ${e_i^I}^*$ are dual to the $e_i$, and  
$e_i^*\cdot e_j=\delta_{ij}$. The left-- and right--moving momenta span a 
Lorentzian even self--dual lattice. The mass formula for the left and 
right--movers is,
$$M_L^2=-c+{{P_L\cdot{P_L}}\over2}+N_L=-1+{{P_R\cdot{P_R}}\over2}+
N_R=M_R^2~,$$
where $N_{L,R}$ are the sum on the left--moving and right--moving oscillators
and $c$ is a normal ordering constant equal to ${1\over2}$ and $0$
for the antiperiodic (NS) and periodic (R) sectors of the NSR fermions \cite{narain}. 
The background fields that produce
the toroidal $SO(12)$ lattice are given by the metric,
\beq
g_{ij}=\left(\matrix{~2&-1& ~0& ~0& ~0& ~0\cr
-1& ~2&-1& ~0& ~0& ~0\cr~0&-1& ~2&-1& ~0& ~0\cr~0& ~0&-1
& ~2&-1&-1\cr ~0& ~0& ~0&-1& ~2& ~0\cr ~0& ~0& ~0&-1& ~0& ~2\cr}\right)~,
\label{gso12}
\eeq
and the antisymmetric tensor,
\beq
b_{ij}=\cases{
g_{ij}&;\ $i>j$,\cr
0&;\ $i=j$,\cr
-g_{ij}&;\ $i<j$.\cr}
\label{bso12}
\eeq
When all the radii of the six-dimensional compactified
manifold are fixed at $R_I=\sqrt2$, it is seen that the
right--moving momenta given by eqs. (\ref{lrmomenta})
produce the root vectors of $SO(12)$ \cite{foc}. 
The next step in the construction of the $Z_2\times Z_2$ orbifold is to add
the two basis vectors $b_1$ and $b_2$ that each breaks $N=4$ space--time supersymmetry
to $N=2$. With a suitable choice of the GSO projection coefficients the
model possesses an ${\rm SO}(4)^3\times {\rm E}_6\times {\rm U}(1)^2
\times {\rm E}_8$ gauge group
and $N=1$ space-time supersymmetry. The matter fields
include 24 generations in the 27 representation of
${\rm E}_6$, eight from each of the sectors $b_1\oplus b_1+\xi_1$,
$b_2\oplus b_2+\xi_1$ and $b_3\oplus b_3+\xi_1$.
Three additional 27 and $\overline{27}$ pairs are obtained
from the Neveu-Schwarz $\oplus~\xi_1$ sector. The same spectrum is obtained by acting with 
the $Z_2\times Z_2$ orbifold on the $SO(12)$ lattice with standard embedding.

A $Z_2\times Z_2$ orbifold at a generic point, however, produces forty--eight fixed points, and hence
forty--eight generations rather than twenty--four. There is therefore a mismatch by a factor
of two between the two models. This mismatch seems puzzling because a priori we do not expect that the
number of fixed points does not depend on the moduli. 

To investigate this issue further we can start with the ${Z}_2\times {Z}_2$ orbifold on
$T_2^1\times T_2^2\times T_2^3$, which gives $(h_{11},h_{21})=(51,3)$.
I will denote the manifold of this model as $X_1$.
We can then add a freely acting twist or shift \cite{befnq} to this model,
which reduces the number of fixed points. Let us first start with the compactified
$T^1_2\times T^2_2\times T^3_2$ torus parameterized by  
three complex coordinates $z_1$, $z_2$ and $z_3$,
with the identification
\beq
z_i=z_i + 1\,, \qquad z_i=z_i+\tau_i \,,
\label{t2cube}
\eeq
where $\tau$ is the complex parameter of each
$T_2$ torus.
With the identification $z_i\rightarrow-z_i$, a single torus
has four fixed points at
\beq
z_i=\{0,{\textstyle{1\over 2}},{\textstyle{1\over 2}}\,\tau,
{\textstyle{1\over 2}} (1+\tau) \}.
\label{fixedtau}
\eeq
With the two ${Z}_2$ twists
\beqn
&& \alpha:(z_1,z_2,z_3)\rightarrow(-z_1,-z_2,~~z_3) \,,
\cr
&&  \beta:(z_1,z_2,z_3)\rightarrow(~~z_1,-z_2,-z_3)\,,
\label{alphabeta}
\eeqn
there are three twisted sectors in this model, $\alpha$,
$\beta$ and $\alpha\beta=\alpha\cdot\beta$, each producing
16 fixed tori, for a total of 48. Adding
to the model generated by the ${Z}_2\times {Z}_2$
twist in (\ref{alphabeta}), the additional shift
\beq
\gamma:(z_1,z_2,z_3)\rightarrow(z_1+{\textstyle{1\over2}},z_2+
{\textstyle{1\over2}},z_3+{\textstyle{1\over2}})
\label{gammashift}
\eeq
produces again fixed tori from the three
twisted sectors $\alpha$, $\beta$ and $\alpha\beta$.
The product of the $\gamma$ shift in (\ref{gammashift})
with any of the twisted sectors does not produce any additional
fixed tori. Therefore, this shift acts freely.
Under the action of the $\gamma$-shift,
the fixed tori from each twisted sector are paired.
Therefore, $\gamma$ reduces
the total number of fixed tori from the twisted sectors   
by a factor of ${2}$,
yielding $(h_{11},h_{21})=(27,3)$. This model therefore
reproduces the data of the ${Z}_2\times {Z}_2$ orbifold
at the free-fermion point in the Narain moduli space.

We note that the freely acting shift (\ref{gammashift}),
added to the ${Z}_2\times {Z}_2$ orbifold at a generic point
of $T_2^1\times T_2^2\times T_2^3$, reproduces the data of the
${Z}_2\times {Z}_2$ orbifold acting on the SO(12) lattice.  
This observation 
does not prove, however, that the vacuum which includes the shift
is identical to the free fermionic model. While the 
massless spectrum of the two models may coincide
their massive excitations, in general, may differ.
The matching of the massive spectra is examined by
constructing the partition function of the ${Z}_2\times {Z}_2$
orbifold of an SO(12) lattice, and subsequently
that of the model at a generic point including the
shift. In effect since the action of the ${Z}_2\times {Z}_2$
orbifold in the two cases is identical the problem
reduces to proving the existence of a freely
acting shift that reproduces the partition function of the
SO(12) lattice
at the free fermionic point. Then since the action of 
the shift and the orbifold projections are commuting
it follows that the two ${Z}_2\times {Z}_2$ orbifolds
are identical. 

The realization of the $SO(12)$ lattice as an orbifold
in achieved by incorporating idenitifications on the 
internal lattice by shift symmetries. It is instructive
for this purpose to study the partition function at a 
generic point in the moduli space, incorporate the 
shifts, and fix the internal radii at the self--dual
point, which then reproduces the partition function
of the $SO(12)$ lattice. The partition function
of the $N=4$ supersymmetric $SO(12)\times E_8\times E_8$
heterotic vacuum is given by
\beq
{Z}=(V_8-S_8)\left[|O_{12}|^2+|V_{12}|^2+|S_{12}|^2+|C_{12}|^2\right]
\left( \bar O_{16} + \bar S_{16}\right) \left( \bar O_{16} + \bar S_{16}
\right) \,, \label{zplus}
\eeq
where ${Z}$ has been written in terms of level-one
${\rm SO} (2n)$ characters (see, for instance, \cite{as})
\beqn
O_{2n} &=& {\textstyle{1\over 2}} \left( {\vartheta_3^n \over \eta^n} +
{\vartheta_4^n \over \eta^n}\right) \,,
\nonumber \\
V_{2n} &=& {\textstyle{1\over 2}} \left( {\vartheta_3^n \over \eta^n} -
{\vartheta_4^n \over \eta^n}\right) \,,
\nonumber \\
S_{2n} &=& {\textstyle{1\over 2}} \left( {\vartheta_2^n \over \eta^n} +
i^{-n} {\vartheta_1^n \over \eta^n} \right) \,,
\nonumber \\
C_{2n} &=& {\textstyle{1\over 2}} \left( {\vartheta_2^n \over \eta^n} -
i^{-n} {\vartheta_1^n \over \eta^n} \right) \,.
\eeqn

On the compact coordinates there are actually three inequivalent ways
in which the shifts
can act. In the more familiar case, they simply translate a generic point 
by half the
length of the circle. As usual, the presence of windings in string 
theory allows shifts on the T-dual circle, or even asymmetric ones, that 
act both on the circle and on its dual. More concretely, for a circle of
length $2 \pi R$, one can have the following options \cite{vwaaf}:
\beqn
A_1\;:&& X_{\rm L,R} \to X_{\rm L,R} + {\textstyle{1\over 2}} \pi R \,,
\nonumber \\
A_2\;:&& X_{\rm L,R} \to X_{\rm L,R} + {\textstyle{1\over 2}} \left(
\pi R \pm {\pi \alpha ' \over R} \right) \,, 
\nonumber \\
A_3\;:&& X_{\rm L,R} \to X_{\rm L,R} \pm {\textstyle{1\over 2}} {\pi \alpha'
\over R} \,.
\label{a1a2a3}
\eeqn
There is, however, a crucial difference among these three choices: while
$A_1$ and $A_3$ shifts can act consistently on any number of coordinates,
level-matching requires instead that the $A_2$-shifts act on (mod) four real 
coordinates. 

Our problem is to find the shift that when acting on the 
lattice $T_2^1\otimes T_2^2\otimes T_2^3$ at a generic point in
the moduli space reproduces the $SO(12)$ lattice when the radii
are fixed at the self--dual point $R=\sqrt{\alpha^\prime}$ \cite{partitions}.
Let us consider for simplicity the case of six orthogonal circles or 
radii $R_i$. The partition function reads
\beq
{Z}_+ = (V_8 - S_8) \, \left( \sum_{m,n} \Lambda_{m,n}
\right)^{\otimes 6}\, \left( \bar O _{16} + \bar S_{16} \right) \left(
\bar O _{16} + \bar S_{16} \right)\,,
\eeq
where as usual, for each circle,
\beq
p_{\rm L,R}^i = {m_i \over R_i} \pm {n_i R_i \over \alpha '} \,,
\eeq
and
\beq
\Lambda_{m,n} = {q^{{\alpha ' \over 4} 
p_{\rm L}^2} \, \bar q ^{{\alpha ' \over 4} p_{\rm R}^2} \over |\eta|^2}\,.
\eeq
We can now act with the ${Z}_2\times {Z}_2$ shifts generated by
\beqn
g\;: & & (A_2 , A_2 ,0 ) \,,
\nonumber \\
h\;: & & (0, A_2 , A_2 ) \,, \label{gfh}
\eeqn
where each $A_2$ acts on a complex coordinate. The resulting partition 
function then reads
\beqn
{Z}_+ &=& {\textstyle{1\over 4}}\, (V_8 - S_8) 
\sum_{m_i , n_i}  \left\{ \left[ 1 + (-1)^{m_1 + m_2 + m_3 + m_4 + n_1 + n_2 +
n_3 + n_4} \right. \right.
\nonumber \\
& & \left. + (-1)^{m_1 + m_2 + m_5 + m_6 + n_1 + n_2 +
n_5 + n_6} + (-1)^{m_3 + m_4 + m_5 + m_6 + n_3 + n_4 +
n_5 + n_6} \right]  
\nonumber \\
& & \left. \times \left( \Lambda_{m_i , n_i}^{1,\ldots ,6} 
+ \Lambda^{1,\ldots,4}_{m_i + {1\over 2}, n_i + {1\over 2}} 
\Lambda^{5,6}_{m_i , n_i} 
+ \Lambda^{1,2,5,6}_{m_i + {1\over 2}, n_i + {1\over 2}} 
\Lambda^{3,4}_{m_i , n_i}
+ \Lambda^{1,2}_{m_i , n_i} 
\Lambda^{3,4,5,6}_{m_i + {1\over 2}, n_i + {1\over 2}} 
\right) \right\}
\nonumber \\
& & \times
\left( \bar O _{16} + \bar S_{16} \right) \left( \bar O_{16} + \bar S_{16}
\right) \label{zpshift}
\eeqn

After some tedious algebra, it is then possible to show that, once evaluated
at the self-dual radius $R_i = \sqrt{\alpha '}$, the 
partition function (\ref{zpshift}) reproduces that at the SO(12) point
(\ref{zplus}). To this end, it suffices to notice that
\beqn
\sum_{m,n} \Lambda_{m,n} (R=\sqrt{\alpha '}) &=& |\chi_0 |^2 + 
|\chi_{1\over 2}
|^2 \,,
\nonumber \\
\sum_{m,n} (-1)^{m+n} \Lambda_{m,n} (R = \sqrt{\alpha '}) &=&
|\chi_0 |^2 - |\chi_{1\over 2} |^2 \,,
\nonumber \\
\sum_{m,n} \Lambda_{m + {1\over 2} , n + {1\over 2}} (R = \sqrt{\alpha '})
&=& \chi_0 \bar\chi_{1\over 2} + \chi_{1\over 2} \bar \chi_0 \,,
\nonumber \\
\sum_{m,n} (-1)^{m+n} \Lambda_{m + {1\over 2} , n + {1\over 2}} 
(R = \sqrt{\alpha '}) &=& \chi_{1\over 2} \bar \chi_0 -
\chi_0 \bar\chi_{1\over 2}  \,,
\eeqn
where
\beqn
\chi_0 &=& \sum_\ell q^{\ell^2} \,,
\nonumber \\
\chi_{1\over 2} &=& \sum_\ell q^{(\ell + {1\over 2})^2} \,,
\eeqn
are the two level-one SU(2) characters, while, standard branching rules,
decompose the SO(12) characters into products of SU(2) ones. For instance,
\beqn
O_{12} &=& \chi_0 \chi_0 \chi_0 \chi_0 \chi_0 \chi_0 +
\chi_0 \chi_0 \chi_{1\over 2} \chi_{1\over 2}
\chi_{1\over 2} \chi_{1\over 2}+
\nonumber \\
& &\chi_{1\over 2} \chi_{1\over 2} \chi_0 
\chi_0 \chi_{1\over 2} \chi_{1\over 2} +
\chi_{1\over 2} \chi_{1\over 2} \chi_{1\over 2} \chi_{1\over 2} 
\chi_0 \chi_0 \,.
\eeqn

The precise form of the orbifold shifts that
produces the $SO(12)$ lattice is given in eq. (\ref{gfh}).  
On the other hand, the shifts given in Eq. (\ref{gammashift}),
and similarly the analogous freely acting shift given by 
$(A_3,A_3,A_3)$, do not reproduce the partition function
of the $SO(12)$ lattice. 
Therefore, the shift in eq. (\ref{gammashift}) does reproduce
the same massless spectrum and symmetries of the ${Z}_2\times {Z}_2$
of the $SO(12)$ lattice, but the partition functions of the 
two models differ! 

Another method to exhibit the reduction of the number of fixed points of
the $Z_2\times Z_2$ orbifold of the $SO(12)$ lattice was presented in
ref. \cite{fft}, using the canonical orbifold method \cite{dhvw}.
The basis vectors of the $SO(12)$ lattice are given by the simple roots
\begin{eqnarray}
e_1 & = & \left( 1 , -1 , 0,0,0,0\right) , \nonumber \\
e_2 & = & \left( 0, 1, -1, 0,0,0\right) , \nonumber \\
e_3 & = & \left( 0 ,0 , 1, -1,0,0\right) , \nonumber \\
e_4 & = & \left( 0, 0, 0, 1,-1,0\right) , \nonumber \\
e_5 & = & \left( 0,0,0,0,1,-1\right) , \nonumber \\
e_6 & = & \left( 0,0,0,0,1,1\right) .\label{eq:so12roots}
\end{eqnarray}
The ${Z}_2 \times {Z}_2$ orbifold action on
a set of six Cartesian coordinates  $x^1, \ldots , x^6$ of the compact
space is specified by: 
\begin{equation}\label{eq:orbiact}
\left( \begin{array}{c}  x^1 \\ \vdots \\ x^6 \end{array}\right)
\rightarrow \theta_1 \left( \begin{array}{c}  x^1 \\ \vdots \\ x^6
\end{array}\right) , \,\,\, \mbox{with} \,\,\, \theta_1 = \left(
\begin{array}{ c c c c c c}
-1     &  0   &   0  &  0  & 0 &  0 \\
0     &  -1   &   0  &  0 &  0 & 0  \\
0     &  0   &   -1 &  0 &  0 & 0 \\
0    &   0   &   0  &  -1 & 0 & 0 \\
0   &    0   &   0  &   0 & 1& 0 \\
0   &   0 &  0 &  0  & 0 & 1  \end{array} \right) 
\end{equation}
and
\begin{equation}\label{eq:orbiact2}
\left( \begin{array}{c}  x^1 \\ \vdots \\ x^6 \end{array}\right)
\rightarrow \theta_2 \left( \begin{array}{c}  x^1 \\ \vdots \\ x^6
\end{array}\right) , \,\,\, \mbox{with} \,\,\, \theta_2 = \left(
\begin{array}{ c c c c c c}
1     &  0   &   0  &  0  & 0 &  0 \\
0     &  1   &   0  &  0 &  0 & 0  \\
0     &  0   &   -1 &  0 &  0 & 0 \\
0    &   0   &   0  &  -1 & 0 & 0 \\
0   &    0   &   0  &   0 & -1& 0 \\
0   &   0 &  0 &  0  & 0 & -1  \end{array} \right)  ,
\end{equation}
where $\theta_1$ and $\theta_2$ are the generators of ${Z}_2\times {Z}_2$.

The orbifold action, by e.g.\ (\ref{eq:orbiact2}), leaves sets of points
invariant, i.e.\ these points differ from their orbifold image by an
SO(12) root lattice shift. For our particular choice of the orbifold
action these sets appear as two dimensional fixed tori. In the
following I will list 16 such two-tori and afterwards argue that some
of these 16 tori are identical. That will leave eight distinct fixed
two-tori. The trivial fixed torus is given as the set
\begin{equation}\label{eq:fix1}
\left\{ \left( x , y , 0,0,0,0\right) \left|\, x,y \in {\mathbb R}^2/
\Lambda^2 \right. \right\} .
\end{equation}
The compactification lattice $\Lambda^2$ is generated by the vectors
$\left( 1 ,1 \right)$ and $\left( 1 , -1 \right)$. This can be
verified by writing
$$ \left( x,y,0,0,0,0\right) = x e_1 + \left( x+y\right) \left(e_2 + e_3 +
e_4 + \frac{1}{2} e_5 + \frac{1}{2} e_6 \right) $$
and identifying minimal shifts in $(x,y)$ shifting the coefficients in
front of lattice vectors by integers.
Now, consider the fixed torus 
\begin{equation}\label{eq:fix2}
\left\{ \left( x , y , 1,0,0,0\right) \left|\,  x,y \in {\mathbb R}^2/
\Lambda^2 \right. \right\} .
\end{equation}
Points on that torus differ from their image point by the lattice
vector $\left( 0,0,2,0,0,0\right)$. The position of the 1
entry can be altered within the last four components by adding SO(12)
root vectors, e.g.\ $\left( 0,0,-1,1,0,0\right)$.  
Next there are fixed tori of the form
\begin{equation}\label{eq:fix3}
\left\{ \left( x , y , \underline{\frac{1}{2},\frac{1}{2},0,0}\right)
  \left|\, x,y \in {\mathbb R}^2/ 
\Lambda^2 \right. \right\} ,
\end{equation}
where the underlined entries can be permuted. Points on these fixed
tori differ from their orbifold image by an SO(12) root, e.g.\ 
$\left( 0 ,0 , 1, 1, 0,0\right)$. There are 
$ \left( \begin{array}{c} 4 \\  2 \end{array} \right) = 6 $
such fixed two-tori. Very similar fixed tori are
\begin{equation}\label{eq:fix4}
\left\{ \left( x , y , \underline{\frac{1}{2},-\frac{1}{2},0,0}\right)
  \left|\, x,y \in {\mathbb R}^2/ 
\Lambda^2 \right. \right\} ,
\end{equation}
where the position of the minus sign can be changed by lattice shifts
( $\left(1/2, -1/2\right) + \left( -1, 1\right) = \left( -1/2,
1/2\right)$).
This yields another set of six fixed tori.
Finally, there are the fixed tori 
\begin{equation}\label{eq:fix5}
\left\{ \left( x , y ,
  \frac{1}{2},\frac{1}{2},\frac{1}{2},\frac{1}{2}\right) 
  \left|\, x,y \in {\mathbb R}^2/ 
\Lambda^2 \right. \right\} 
\end{equation}
and
\begin{equation}\label{eq:fix6}
\left\{ \left( x , y ,
  \frac{1}{2},\frac{1}{2},\frac{1}{2},-\frac{1}{2}\right) 
  \left|\, x,y \in {\mathbb R}^2/ 
\Lambda^2 \right. \right\} .
\end{equation}
So, altogether there are 16 fixed tori. Some of these are equivalent.
Consider the fixed torus
(\ref{eq:fix2}) and add the SO(12) root vector $\left(
1,0,-1,0,0,0\right)$. This yields an equivalent expression for
(\ref{eq:fix2}) 
\begin{equation}
\left\{ \left( x+1 , y ,
  0,0,0,0\right) 
  \left|\, x,y \in {\mathbb R}^2/ 
\Lambda^2 \right. \right\} .
\end{equation}
But this is the same fixed torus as (\ref{eq:fix1}), merely the origin
for the $x$ coordinate has been shifted by one. Similar arguments show
that the tori in (\ref{eq:fix3}) and (\ref{eq:fix4}) as well as the
tori (\ref{eq:fix5}) and (\ref{eq:fix6}) are mutually equivalent. So,
finally we are left with eight inequivalent fixed tori.

For the ${Z}_2 \times {Z}_2$ orbifold we add another
${Z}_2$ action $\theta_1$ (\ref{eq:orbiact}).
For this ${Z}_2 \times {Z}_2$ action we obtain eight fixed
tori under the action of $\theta_1$, eight fixed tori under the action
of $\theta_2$ 
and eight fixed tori under the action of $\theta_1\theta_2$. Hence, the
total number 
of fixed tori is 24. On each fixed tori, labeled by a complex
coordinate $z_i$ $i=1,2,3$, there is a non--trivial identification 
imposed by the second $Z_2$ orbifold $z_i\rightarrow -z_i$.
The results of this identification is that the fixed torus 
degenerates to $P_1$. This degeneration is the origin of the 
chirality in this construction \cite{df2}.

The model discussed above represent an explicit case in which the
correspondence between the free fermion construction and the bosonic
construction has been explicitly demonstrated at the level of the string
partition function, {\it i.e.} at the massless as well as the massive
string spectrum. Of course, there are many more vacua that can be constructed.
The classification of symmetric $Z_2\times Z_2$ orbifolds with standard 
embedding using bosonic techniques was studied in \cite{df2,dw}. In ref. 
\cite{classification} a more general classification using fermionic 
techniques was presented. In this classification the gauge degrees 
of freedom are separated into four modular blocks, which is the 
most general separation compatible with modular invariance. The
space of vacua spanned includes models with $(2,2)$ world--sheet
supersymmetry as well as models in which the right--moving $N=2$ 
world--sheet supersymmetry is broken. Hence, this analysis also
includes vacua with non--standard embedding, and revealed a
spinor--vector duality map in the space of these vacua
\cite{classification}. The two dimensional Fermi--Bose equivalence,
however, entails that every model constructed using the fermionic
techniques can also be constructed using the bosonic techniques.
Deriving this dictionary will provide further insight into 
the properties of the string vacua. 

\ni {\bf Acknowledgements}

I would like to thank the organisers for the opportunity to speak at the
6$^{th}$ Simons workshop; the University of Oxford for hospitality
while this work was written; and my collaborators over the years, with
special thanks to Costas Kounnas and John Rizos. This work is supported
in part by the STFC under contract PP/D000416/1 and by the 
EU UniverseNet network under contract MRTN--CT--2006--035863--1.


\begin{thebibliography}{99}

\bibitem{fsu5} I.\ Antoniadis, J.\ Ellis, J.\ Hagelin and D.V.\ Nanopoulos
                \PLB{231}{1989}{65}.

\bibitem{fny} A.E.\ Faraggi, D.V.\ Nanopoulos and K.\ Yuan,
                                                 \NPB{335}{1990}{347};\\
                \AEF, \PRD{46}{1992}{3204}.

\bibitem{alr}	I. Antoniadis. G.K. Leontaris and J. Rizos,
                                \PLB{245}{1990}{161}.

\bibitem{nahe} A.E. Faraggi and D.V. Nanopoulos, \PRD{48}{1993}{3288}.

\bibitem{eu} \AEF, \PLB{278}{1992}{131}; \NPB{403}{1993}{101}.

\bibitem{top} \AEF, \PLB{274}{1992}{47}; \PLB{339}{1994}{223}.

\bibitem{cslm}	A.E. Faraggi, \NPB{387}{1992}{239}.

\bibitem{otherrsm} S. Chaudhuri \etal, \NPB{469}{1996}{357};\\
		   G.B. Cleaver \etal, \NPB{525}{1998}{3};
                                \NPB{545}{1998}{47}.

\bibitem{cfn} G.B.\ Cleaver, \AEF~ and D.V.\ Nanopoulos,
                       \PLB{455}{1999}{135}; \IJMP{16}{2001}{425};\\
              G.B.\ Cleaver, \AEF, D.V.\ Nanopoulos and J.W.\ Walker,
                                       \MPLA{15}{2000}{1191};                   
                                        \NPB{593}{2001}{471};
                                        \NPB{620}{2002}{259};\\
                                        G.\ Cleaver, \IJMPA{16S1C}{2001}{949}.

\bibitem{cfs} G.B.\ Cleaver, A.E.\ Faraggi and C.\ Savage,
                                \PRD{63}{2001}{066001};\\
              G.B.\ Cleaver, D.J.\ Clements and A.E.\ Faraggi,
                        \PRD{65}{2002}{106003}.
 
\bibitem{su421} G. Cleaver, A.E. Faraggi and S.~Nooij, \NPB{672}{2003}{64}. 

\bibitem{fmt} \AEF, E. Manno and C. Timirgaziu, \EPJC{50}{2007}{701};\\
   G. Cleaver, \AEF, E. Manno and C. Timirgaziu, \PRD{78}{2008}{046009}.

\bibitem{fff} 
  I.~Antoniadis, C.~P. Bachas, and C.~Kounnas, \NPB{289}{1987}{87};\\
  H. Kawai, D.C. Lewellen, and S.H.-H. Tye, \NPB{288}{1987}{1};\\
  I.~Antoniadis and C.~P. Bachas, \NPB{289}{1987}{87}.

\bibitem{egrs} S. Elitzur, E. Gross, E. Rabinovici and N. Seiberg,
  \NPB{283}{1987}{413}.

\bibitem{gepner} D. Gepner, \PLB{199}{1987}{380}.

\bibitem{top95} \AEF, \PLB{377}{1996}{43}; \NPB{487}{1997}{55}.

\bibitem{fmm} \AEF, \NPB{407}{1993}{57}.

\bibitem{ckm} I. Antoniadis, J. Rizos and K. Tamvakis, \PLB{278}{1992}{257};\\
             \AEF~and E. Halyo, \PLB{307}{1993}{305}; \NPB{416}{1994}{63}.

\bibitem{seesaw} I. Antoniadis, J. Rizos and K. Tamvakis, 
					\PLB{279}{1992}{281};\\
                 \AEF~and E. Halyo, \PLB{307}{1993}{311}.

\bibitem{seesawII} C. Coriano and \AEF, \PLB{581}{2004}{99}.

\bibitem{gcu} I. Antoniadis, J. Ellis, R. Lacaze and D.V. Nanopoulos, 
			\PLB{268}{1991}{188};\\
              \AEF, \PLB{302}{1992}{202}. 

\bibitem{df} K.R. Dienes and \AEF, \PRL{75}{1995}{2646};
                                           \NPB{457}{1995}{409}.             
\bibitem{ps} \AEF, \NPB{428}{1994}{111}; 
                   \PLB{499}{2001}{147};
                   \PLB{520}{2001}{337};\\
             J.C. Pati, \PLB{388}{1996}{532}.

\bibitem{fp2} I. Antoniadis, J. Ellis, A. Lahanas and D.V. Nanopoulos,
						\PLB{241}{1990}{24};\\
              \AEF~ and J.C. Pati, \NPB{526}{1998}{21};\\
              \AEF~ and O. Vives, \NPB{641}{2002}{93}.

\bibitem{modulifixing} \AEF, \NPB{728}{2005}{83}. 

\bibitem{classification} \AEF, C. Kounnas, S.E.M. Nooij and J. Rizos, 
                               \NPB{695}{2004}{41};\\ 
                               A.E.Faraggi, C.~Kounnas and J.~Rizos,
                               \PLB{648}{2007}{84}; \NPB{774}{2007}{208};
                               \NPB{799}{2008}{19};\\
          T. Catelin-Jullien, \AEF, C. Kounnas and J. Rizos, arXiv:0807.4084.

\bibitem{zp} \AEF~and D.V. Nanopoulos, \MODA{6}{1991}{61};\\
 C. Coriano, \AEF~and M. Guzzi, \EPJC{53}{2008}{421}; \PRD{78}{2008}{015012}.

\bibitem{ccf}   S. Chang, C. Coriano and \AEF,  
                                \NPB{477}{1996}{65};\\
	C. Coriano, \AEF~and M. Plumacher, \NPB{614}{2001}{233}.

\bibitem{dedes} A. Dedes and \AEF, \PRD{62}{2000}{016010}.

\bibitem{sdvs} \AEF, \IJMP{19}{2004}{5523}; hep-th/0411118.

\bibitem{dw} R. Donagi and K. Wendland, arXiv:0808.0330. 

\bibitem{narain} K.S. Narain, \PLB{169}{1986}{41};\\
                 K.S. Narain, M.H. Sarmadi and E. Witten, 
				\NPB{279}{1987}{369}.

\bibitem{foc} \AEF, \PLB{326}{1994}{62}.

\bibitem{befnq} P. Berglund \etal, \PLB{433}{1998}{269};
				   \IJMP{15}{2000}{1345}.

\bibitem{as} See {\it e.g.}: C. Angelantonj and A. Sagnotti, 
			\PRT{371}{2002}{1}.

\bibitem{vwaaf} C. Vafa and E. Witten,  1996
	{\it Nucl.Phys.Proc.Suppl.} {\bf46} (1996) 225;\\
	C. Angelantonj, I. Antoniadis and K. F\"orger, \NPB{555}{1999}{116}.

\bibitem{partitions} \AEF, \PLB{544}{2002}{207}.

\bibitem{dhvw} L. Dixon, J.A. Harvey, C. Vafa and E. Witten,
  \NPB{274}{1986}{285}.

\bibitem{fft} \AEF, S. Forste and C. Timirgaziu, \JHEP{0608}{2006}{057}.

\bibitem{df2} R. Donagi and \AEF, \NPB{694}{2004}{187}.


\end{thebibliography}
 \end{document}